\documentclass[12pt,osajnl2,preprint,showpacs]{revtex4}
\usepackage[draft]{hyperref} 
\DeclareRobustCommand{\baselinestretch{2.2}}

\usepackage{graphicx,natbib}

\begin{document}

\title{Optimal Dark Hole Generation via Two Deformable Mirrors with Stroke Minimization}

\author{Laurent Pueyo}
 \affiliation{Jet Propulsion Laboratory, California Insitute of Technology, 4800 Oak Grove Drive, Pasadena, CA, 91109, USA}
\email{lpueyo@jpl.nasa.gov}

\author{Jason Kay}
 \affiliation{Mechanical and Aerospace Engineering, Princeton University, Princeton, NJ, 08544, USA}

\author{N. Jeremy Kasdin}
 \affiliation{Mechanical and Aerospace Engineering, Princeton University, Princeton, NJ, 08544, USA}

\author{Tyler Groff}
 \affiliation{Mechanical and Aerospace Engineering, Princeton University, Princeton, NJ, 08544, USA}

\author{Michael Mc Elwain}
 \affiliation{Mechanical and Aerospace Engineering, Princeton University, Princeton, NJ, 08544, USA}

\author{Amir Give'on}
 \affiliation{Jet Propulsion Laboratory, California Insitute of Technology, 4800 Oak Grove Drive, Pasadena CA, 91109, USA}
 
 \author{Ruslan Belikov}
 \affiliation{NASA Ames Research Center, Moffett Field, California 94035  USA}



\begin{abstract}
The past decade has seen a significant growth in research targeted at space based observatories for imaging exo-solar planets. The challenge is in designing an imaging system for high-contrast. Even with a perfect coronagraph that modifies the point spread function to achieve high-contrast, wavefront sensing and control is needed to correct the errors in the optics and generate a ``dark hole''. The high-contrast imaging laboratory at Princeton University is equipped with two Boston Micromachines Kilo-DMs. We review here an algorithm designed to achieve high-contrast on both sides of the image plane while minimizing the stroke necessary from each deformable mirror (DM). This algorithm uses the first DM to correct for amplitude aberrations and the second DM to create a  flat wavefront in the pupil plane. We then show the first results obtained at Princeton with this correction algorithm, and we demonstrate a symmetric dark hole in monochromatic light.
\end{abstract}

\maketitle 

\section{Introduction}
\label{sec:introduction}

With the need to image faint exo-planets close to their parent star, wavefront control for space-based high contrast imaging has received tremendous attention over the past few years. The idea of completely canceling starlight using coherent subtraction via a Deformable Mirror (DM) was first introduced by \citet{1995PASP..107..386M}, where they proved, using non linear numerical tools, the theoretical feasibility of a very high contrast Dark Hole in the image plane of a telescope. Experimentally the first idea tested was based on a linearization of this previous result:  cycle through a set of arbitrary DM configurations and choose the one yielding the best contrast in an algorithm called ``Speckle Nulling'' (e.g., \citet{BrownBurrows}, \citet{TraugerProof}). \citet{2006ApJ...638..488B} proposed a refinement of this solution that  yielded much faster convergence rates: they separated the estimation and the correction stages and reduced the second one to a simple matrix inversion based on an energy minimization criteria. However this method presented numerical caveats when associated to a coronagraph since it required the inversion of an ill-conditioned matrix. \citet{2007AAS...21113520G} proposed a solution to regularize the problem  based on Electrical Field Conjugation. In this paper we introduce an alternative correction method that fully solves the non-linear problem while using the smallest DM deformation possible. We not only present the theory underlying these algorithms but also provide an experimental validation of the Stroke Minimization method using one and two DMs. Previous experimental results of high contrast Dark Hole formation only achieved high contrast in one-half of the image plane.  Because of the unique dual DM feature of the Princeton University High Contrast Laboratory, this paper is the first experimental report of symmetric high contrast point spread functions.

One of the characteristics of high contrast wavefront control in space is focal plane wavefront sensing.  Using the science camera to also estimate the wavefront avoids potential non-common path errors from conventional pupil estimation.  The control algorithms discussed in this paper, therefore, all assume perfect estimates of the field in the region of the image plane where we seek high contrast.  The experiment uses an estimation algorithm that retrieves the wavefront based on random DM diversity  (\citet{AmirPeekaBoo} and \citet{2007AAS...21113520G}). That is, probes are set on the DM in order to cause interference with the aberrated field. A careful analysis of the resulting speckles lead to the retrieval of the complex field.  

For all the analysis and simulation of the various control approaches in this paper we assume perfect estimates of the focal plane field.  

\section{The Optimal Dark Hole Problem}

The goal of the correction problem is to cancel the starlight due to optical imperfection so that the contrast in the image plane is the one for which the coronagraph was designed.  It was first shown in \citet{1995PASP..107..386M} that this could be formulated as a nonlinear optimization problem.  While the problem itself is straightforward to state, many solution approaches are possible.  Here, we follow the notation of  \citet{2007AAS...21113520G}.  We model the coronagraph as a general linear transformation, $C$, between the electric field at the deformable mirror, $E_0$, and the final electric field at the image plane of the camera,
\begin{equation}
E_f = C\{ E_0 \}.
\label{eq:e_f}
\end{equation}
When the DM is at the pupil, the linear transformation $C$ is a Fourier transform. When the DM is not at the pupil, the operator includes Fresnel propagations from the DM to the pupil. The linear mapping could represent any number of possible coronagraphs (shaped pupils, Lyot type coronagraphs, pupil mapping, etc.).  In most cases, it will simply be a series of Fourier transforms and convolutions involving various mask types.  For the simulations in this paper, we assume a shaped pupil coronagraph because of its simplicity.  There, $C\{E_0\}$ simply represents the Fourier transform of the field at the shaped pupil.  For this analysis, the DM is assumed to be in a plane conjugate with the shaped pupil, while the experimental results presented in later sections, take into account propagation between the DMs and the plane of the shaped pupil.

If we include amplitude and phase aberrations and a single DM to correct,  the electric field at the DM can be written as (\citet{2007AAS...21113520G}),
\begin{equation}
E_0 = A(x,y) e^{\alpha_\lambda(x,y) + i\frac{2\pi}{\lambda} \beta_\lambda(x,y)}e^{i\frac{2\pi}{\lambda}\psi(x,y)},
\label{eq:aberrated_pupil}
\end{equation}
where $\alpha_\lambda(x,y)$ and $\beta_\lambda(x,y)$ are respectively the amplitude and phase aberrations across the pupil (and may differ with wavelength due to propagation effects), $A(x,y)$ is the pupil apodization, and $\psi(x,y)$ is the DM height in units of wavefront. The problem is to find a DM surface to sufficiently cancel the aberrations and restore the contrast in some region of the image plane, often called the ``dark hole''.

This is an infinite-dimensional problem and cannot be solved exactly.  One approach to making it tractable is to approximate the DM surface height by a finite sum of basis functions,
\begin{equation}
\psi(x,y) = \lambda_0 \sum_{k=1}^N a_{k} f_{k}(x,y).
\end{equation}
where $\lambda_0$ is the central wavelength in the band considered. The problem is now to choose the coefficients $a_{k}$ to achieve the desired performance.  The basis functions can be any convenient set, such as Legendre polynomials, Zernike's, Fourier modes, Chebyshev polynomials, or others.  Depending upon the choice of basis function, the coefficients, $a_{k}$, will be some function of the voltages applied to the DM.  Selecting a particular expansion is a tradeoff among various objectives.  One objective might be to minimize the number of terms for an adequate fit.  Another might be to simplify the relationship between the coefficients and the voltages.  Most often, the $f_{k}$ are chosen to be the so-called {\em influence functions}.  These are the shapes the DM takes when a voltage is applied to only a single actuator. The assumption is then made that superposition holds (that is, that an arbitrary shape can be found by summing influence functions).  For this choice, the coefficients have a one-to-one correspondence to the voltage applied to the $k^{th}$ actuator.  The experimental work reported in this paper uses such an influence function basis. 

To further simplify our notation, we write the coefficients to be found as a single column matrix, $X = \left [ a_1, a_{2},\ldots,a_{N-1},a_{N}\right ]^T$ and the basis functions in the row matrix $F(x,y) = \left [ f_1(x,y), f_2(x,y), \ldots, f_{N-1}(x,y), f_{N}(x,y) \right ]$, which lets us write the DM surface in matrix form,
\begin{equation}
\psi(x,y) = \lambda_0 F(x,y) X.
\end{equation}
Finally, for convenience we streamline our notation by replacing the complex exponentials in Eq. (\ref{eq:aberrated_pupil}) by a function $h$,
\begin{equation}
E_0 = A(x,y)h(\gamma_\lambda(x,y),X)
\label{eq:E0_h}
\end{equation}
where $\gamma_\lambda = \alpha + i\frac{2\pi}{\lambda}\beta$ is the complex, wavelength dependent abberation and $h$ represents the nonlinear dependence on the aberration and DM settings.  

The simplest correction algorithm would be to choose the coefficients such that the DM surface cancels, or {\em conjugates}, the aberrations in Eq. (\ref{eq:aberrated_pupil}).  This is the typical approach used in conventional, ground based adaptive optics (AO).  We notice immediately two limitations.  First, with a single DM it is only possible to correct the field at a single wavelength.  Second, since the DM is a phase correcting device, it nominally only corrects for the phase aberration.  However, 
complete phase conjugation across the pupil is more than is needed, as we seek a dark hole in only a limited region of the image plane.  In fact, we show in the appendix that by correcting only the part of $\alpha(x,y)$ whose image plane electrical field is centro-symmetric with respect to the optics axis , it is possible to correct for both amplitude and phase in a dark hole on only one half of the image plane with a single DM.  For symmetric dark holes, two DMs are necessary.  In \citet{2007ApJ...666..609P}, we show how, in principle, two DMs can be used to correct for both amplitude and phase across the image plane and at multiple wavelengths.  We discuss two DM corrections in Section \ref{sec:two_DMs}.

Deterministic phase conjugation also requires knowledge of $\beta_\lambda(x,y)$.  In classical AO, this is done via a wavefront sensor in a diverted beam.  For extreme high contrast in space, however, non-common path errors result in an uncorrectable component that degrades the desired contrast.  Our goal is to develop algorithms for correction using only measurements of the image plane field, avoiding the need to backpropagate to estimates of the pupil aberrations.
In the remainder of this section we review past approaches and present a new approach, {\em stroke minimization}.  We also continue to restrict ourselves to algorithms incorporating one DM at a single wavelength.  In section \ref{sec:two_DMs} we show how to generalize this one DM solution to two sequential DMs and present the first experimental validation of two DM dark hole generation. 
\subsection{Energy Minimization}
One approach to achieving a dark hole is to minimize the energy in the desired region of the image plane as first proposed by \citet{1995PASP..107..386M}.  While this first form of the problem resulted in a complicated non-linear optimization, \citet{2006ApJ...638..488B} presented a simplified, linear form which we review in this section.

The total energy  in some region $\mathcal S$ of the image plane is written as the integral of the intensity over that region,
\begin{equation}
\mathcal{E}_\mathcal{S} = \left < E_f, E_f \right >_\mathcal{S} = \int \int_\mathcal{S} E_f E_f^* d\xi d\eta,
\label{eq:dark_hole_energy}
\end{equation}
where $^*$ represents the complex conjugate and $(\xi,\eta)$ are image plane coordinates and $E_f = C \{ E_0 \}$.

In Energy Minimization we seek to minimize $\mathcal{E}_\mathcal{S}$ over $X$.  There are a number of ways one might do this.  For example, if we let $g(X) = \left < E_f, E_f \right >_\mathcal{S}$, then a simple steepest descent algorithm gives,
\begin{equation}
X_{n+1} = X_n - \nu \nabla_X g(X_n)
\end{equation}
for some constant $\nu$, where  $\nabla_X g(X_n)$ is the gradient of  $g(X)$ evaluated at $X_n$. While simple to formulate, steepest descent is known to take many step to converge down a long narrow valley and tends to overshoot farther than the minimum (\citet{NumRecip}).  Instead, we might choose a Newton algorithm based on a Taylor series expansion of $g(X)$,
\begin{equation}
X_{n+1} = X_n - H_n^{-1} \nabla_X g(X_n)
\end{equation}
 where $H_n$ is the hessian matrix of $g(X)$ at $X=X_n$.  However, it is a common problem with Newton's method that when far from the solution, the Hessian can become negative definite and the step no longer descends.  The usual approach is to use Gauss' correction, resulting in the Gauss-Newton iteration,
\begin{equation}
X_{n+1} = X_n - ( \nabla_X g(X_n)^T  \nabla_X g(X_n))^{-1}  \nabla_X g(X_n)^T.
\end{equation}
The correction step to the coefficients is given by the pseudo-inverse of the Jacobian of the total energy in the dark hole.
 
Using the expression for the electric field in Eqs. \ref{eq:e_f} and \ref{eq:aberrated_pupil}, the Jacobian can be written as:
\begin{eqnarray}
 \nabla_X g(X_n) &=&  2* \Re \left[ \left < C\{E_0(X_n)\}, i \frac{2 \pi \lambda_0}{\lambda} C\{E_0(X_n) F\}  \right >_\mathcal{S}  \right]
\end{eqnarray}
where $\Re$ stands for the real part of a complex number. \citet{2006ApJ...638..488B}'s insight was realizing that computing the gradient in the Gauss approximation to the Hessian around $X=0$, and neglecting cross terms between $\gamma_{\lambda}(x,y)$ and $\psi$, would greatly simplify the problem at a small cost in convergence rate. Following their presentation, we write a first order expansion of $h(\gamma_\lambda(x,y),X)$ as
\begin{equation}
h(\gamma_\lambda(x,y),\delta X_n) \simeq 1+ \gamma_{\lambda}(x,y) +  i \frac{2 \pi \lambda_0}{\lambda} F(x,y) \delta X_n
\label{EqLinExp}
\end{equation}
where we have expanded about $X = 0$ and ignored the cross term in $i \frac{2 \pi \lambda_0}{\lambda} F(x,y) \delta X_n \; \gamma_{\lambda}(x,y) $. Ignoring this cross term, leads to an errors of about $| \gamma_{\lambda}|$ percent in the actual gradient. This error in theory slows down the convergence rate. However, as discussed later in this section, for this application such an error in the gradient is not the convergence rate limiting factor. In any case, when the starting wavefront error is small, starting contrast below $10^{-4}$, the error on the gradient does not impact the minima towards which the algorithm converges. As a consequence we can write,
\begin{eqnarray}
 \nabla_X g(X_n) & \simeq & 2* \Re \left[ \left < E_f,  i \frac{2 \pi \lambda_0}{\lambda} C \{A F \}  \right >_\mathcal{S}  \right] \\
  \nabla_X g(X_n)^T   \nabla_X g(X_n) & \simeq & - 2 (2 \frac{2 \pi \lambda_0}{\lambda})^2 * \Re \left[ \left <C \{A F \} ,  C \{A F \}  \right >_\mathcal{S}  \right].
\end{eqnarray}
This approximation significantly simplifies the algorithm as $ \nabla_X g(X_n)^T   \nabla_X g(X_n) $ can now be pre-computed and $ \nabla_X g(X_n)$ only depends on the projection of the image plane field at iteration $n$ on the modal matrix $C\{ A F(x,y) \}$.

Implementing energy minimization thus requires only an estimate of the electric field at the image plane, $E_f$.  \citet{2006ApJ...638..488B}  suggested finding this field via image plane measurements only and a DM diversity estimation scheme, avoiding any non-common path errors associated with pupil sensors. They devised a reconstructor for $E_f$ that used several DM settings to disentangle the ambiguity of intensity measurements at the science camera. 

Unfortunately, a common problem with the Gauss-Newton method is that $ \nabla_X g(X_n)^T  \nabla_X g(X_n)$ can drop rank.  In fact, this is a signficant problem for dark hole generation as the image plane area being minimized is small enough that  $X_n$ is underdetermined.  One common solution is the Levenberg algorithm, where a constant ``damping parameter" is added to the search direction,
\begin{equation}
X_{n+1} = X_n - ( \nabla_X g(X_n)^T  \nabla_X g(X_n)+\mu \; I)^{-1}  \nabla_X g(X_n)^T.
\end{equation}
where $\mu$ is a regularization parameter and $I$ the identity matrix. This approach provides a balance between Gauss-Newton and steepest descent, allowing faster convergence when far away and slower convergence (avoiding overshoot) as the minimum is approached.  It also guarantees invertibility of the weighting matrix.  Marquardt suggested a modification, replacing the identity matrix with the square of the Gauss approximation to the Hessian,
\begin{eqnarray}
X_{n+1} &=& X_n - (\nabla_X g(X_n)^T  \nabla_X g(X_n)  \nonumber \\
& & +\mu \; \mbox{diag}((\nabla_X g(X_n)^T  \nabla_X g(X_n))^T (\nabla_X g(X_n)^T  \nabla_X g(X_n))))^{-1}  \nabla_X g(X_n)^T.
\end{eqnarray}
This is known as the Levenberg-Marquardt algorithm.  Marquardt's damping correction is very similar to Tikhonov regularization in ill-posed linear problems.  \citet{1995PASP..107..386M} used this algorithm in a computationally expensive fashion since they were evaluating  the gradient $ \nabla_X g(X_n)$ numerically  and also finding $\nabla_X g(X_n)^T  \nabla_X g(X_n)$ at each iteration. The slower convergence rate due to the  approximation in Eq.~\ref{EqLinExp} is a minor drawback compared to the efficiencies gained by computing the approximate Hessian only.

In practice, if it were possible to measure $E_f$ perfectly with a precision better than $10^{-5}$, then only one series of numerical iteration would be sufficient to converge to a solution that would yield a contrast when the magnitude square of the errors in the estimate is below $10^{-10}$. However, in the presence of photon and camera noise, such a precision will never be achieved and a new estimate is needed each time the contrast improves.  Moreover, since the estimation is based on DM diversity,  and the actual DM shape is not fully known, the estimate of $E_f$ becomes biased.  As a consequence, while in theory the algorithm presented here could only consist of numerical iterations with only one measurement of $E_f$, in reality it needs to include iterations with successive estimates of the field due to experimental limitations.

\subsection{Electric Field Conjugation}\label{sec_EFC}

While some of the numerical difficulties of energy minimization could be alleviated by trying other quasi-Newton search techniques, such a path is not fruitfull.  Since we are minimizing the energy, we are not necessarily guaranteeing  the contrast we are trying to achieve.  That is, the line search approach never uses the fact that a coronagraph has been implemented to achieve high contrast.  One alternative approach is the {\em electric field conjugation} (EFC) algorithm introduced by \citet{2007AAS...21113520G}.  In this algorithm, Give'on replaced the optimization with a root finding problem.  Since we have a desired field in the image plane, $E_D(\xi,\eta) = C\{A(x,y)\}$, the control problem becomes finding the DM settings required to achieve that field in $\mathcal{S}$,
\begin{equation}
E_f - E_D = 0.
\end{equation}
if we use a continuous version of the image plane, this problem is infinite dimensional and nonlinear in the $a_k$.  It is made tractable by discretizing the field at a finite number of points in the image plane.  Give'on solved for the roots by assuming the DM surface height to be small and expanding the complex exponential in a Taylor series,
\begin{equation}
e^{i\frac{2\pi}{\lambda}\psi} \cong e^{i\frac{2\pi}{\lambda}\psi_n} \left (1 + \frac{i 2 \pi}{\lambda}  \delta \psi_{n+1} \right )
\end{equation}
where we have written the DM surface as the surface at the previous iterate, $\psi_n$, plus a small correction, $\delta \psi_{n+1}$.

This lets us write the root finding problem at each iteration as a simple linear equation,
\begin{equation}
\tilde E_n + i \frac{2\pi \lambda_0}{\lambda}  C\{Ah(\gamma_{\lambda},X_n) F\} \delta X_{n+1} = 0
\label{eq:EFC}
\end{equation}
where $\tilde E_n = E_D - C\{Ae^{\alpha_{\lambda} + \frac{i2\pi}{\lambda}\beta_{\lambda}  + \frac{i2\pi}{\lambda}\psi_n}\} = E_D - (E_f)_n$ is the error between the desired and previously corrected aberrated electric field at the image plane and $\delta X_{n+1}$ is the change in the DM coefficients $X_n$ at iteration $n+1$.  Since in all of the algorithms we are considering, we assume only an estimate of the electric field in $\mathcal{S}$, we can further simplify by expanding $C\{Ah_nf_k\}$ and approximating it by it's zero order term,
\begin{equation}
\tilde E_n +  i \frac{2\pi \lambda_0}{\lambda}  C\{A F\} \delta X_{n+1}= 0.
\label{eq:EFC_simple}
\end{equation}

This is a linear equation in $\delta X_{n+1}$ and can thus be solved for the DM increment given a previous measurement of the image plane field in $\mathcal{S}$ at step $n$, $C\{Ae^{\alpha + \frac{i2\pi}{\lambda}\beta + \frac{i2\pi}{\lambda}\psi_n}\}$.   
 
However, for it to be invertible,  the same number of points in the image plane must be taken as there are coefficients, $a_k$.  Unfortunately, measurements are only available at the pixel spacings in the dark hole, which typically are much fewer in number than the available actuators.  The result is an under-determined problem for the $\delta X_{n+1}$.  The approach taken in \citet{2007AAS...21113520G} is to form the pseudo-inverse of $C\{AF\}$, which is equivalent to finding the minimum norm solution for $\delta X_{n+1}$.  Unfortunately, this too can suffer from numerical difficulties.  In particular, for a given coronagraph there is no guarantee that even the pseudo-inverse of $C\{AF\}$ is  well-behaved.  This is particularly true for shaped pupils since many of the DM actuators are covered by the opaque regions of the mask.

\subsection{Stroke Minimization}
We solve many of these problems in our new approach, which we term ``stroke minimization''.  In stroke minimization, we eliminate the dimensionality problem by minimizing the finite number of coefficients  in the DM surface expansion rather than the field itself.  This has the effect of minimizing the average stroke of the DM actuators, an important criteria for the small devices being used.  We include in the minimization the constraint that the field meet the contrast requirement in the dark hole.  

The optimization problem is thus written,
\begin{eqnarray*}
\mbox{minimize} & & \frac{1}{2} \sum_{k=1}^N a_k^2 \\
\mbox{subject to} & & \mathcal{E}_{\mathcal{S}} \le 10^{-\mathcal{C}}
\end{eqnarray*}
where $\mathcal{E}_{\mathcal{S}} = \left <E_f,E_f\right >_\mathcal{S}$ is the integrated intensity in the dark hole given by Eq. (\ref{eq:dark_hole_energy}) and $C$ is the contrast desired.

While we have solved the dimensionality problem, this is still a difficult nonlinear program.  A common solution approach is to expand the constraint in a Taylor series about the previous coefficient settings and replace the optimization with a quadratic subprogram at each step.  Returning to our matrix notation where $X_n$ is the matrix of coefficients at step $n$ and $\delta X_{n+1}$ is the change in actuator settings, we can approximate the field at the DM, $E_0$, in Eq. (\ref{eq:E0_h}),
\begin{equation}
E_0 \cong A(x,y) \left [ h_n(\gamma_\lambda(x,y),X_n) + \left . \frac{\partial h}{\partial X}\right |_{X_n} \delta X_{n+1} \right ]
\end{equation}
where the Jacobian of $h$ is the row matrix,
\begin{equation}
J_n = \left . \frac{\partial h}{\partial X}\right |_{X_n} = i \frac{2 \pi\lambda_0}{\lambda} e^{\alpha_{\lambda}(x,y) + i\frac{2\pi}{\lambda} \beta_{\lambda}(x,y)}e^{i\frac{2\pi}{\lambda}\psi_n(x,y)}F(x,y).
\end{equation}

The integrated intensity can now be approximated using this first order expansion,
\begin{equation}
\mathcal{E}_\mathcal{S} \cong \int \int_{\mathcal{S}(\xi,\eta)} \left [C\{Ah_n\} + C\{AJ_n\} \delta X_{n+1}\right ]^* \left [C\{Ah_n\} + C\{AJ_n\} \delta X_{n+1}\right ] d\xi d\eta.
\end{equation}
Multiplying through gives us,
\begin{eqnarray}
\mathcal{E}_\mathcal{S} &\cong& \underbrace{\int \int_{\mathcal{S}} C\{Ah_n\}^*C\{Ah_n\} d\xi d\eta}_{(\mathcal{E}_\mathcal{S})_n} + 
2 \Re \left \{ \int \int_{\mathcal{S}} C\{Ah_n\}^* C\{AJ_n\} \delta X_{n+1} d\xi d\eta \right \} \nonumber \\
& & + \int \int_{\mathcal{S}} \delta X_{n+1} ^T C\{AJ_n\}^* C\{AJ_n\}^T \delta X_{n+1} d\xi d\eta \nonumber
\end{eqnarray}
where we have noted that the first term is just the measured energy in the dark hole at iteration $n$.  This lets us write the quadratic subprogram,
\begin{eqnarray}
\mbox{minimize} & & \frac{1}{2} (X_n + \delta X_{n+1})^T W^{-1} (X_n + \delta X_{n+1} ) \nonumber \\
\mbox{subject to} & & \delta X_{n+1}^T M_n \delta X_{n+1} + B_n \delta X_{n+1} +d_n \le 10^{-\mathcal{C}} \nonumber
\end{eqnarray}
where we have generalized to allow for an arbitrary weighting among the coefficients, through the matrix $W$, and,
\begin{eqnarray*}
d_n & = & \int \int_{\mathcal{S}} C\{Ah_n\}^*C\{Ah_n\} d\xi d\eta \\
B_n & = & 2\Re \left \{ \int \int_{\mathcal{S}} C\{Ah_n\}^* C\{AJ_n\} d\xi d\eta \right \} \\
M_n & = & \int \int_{\mathcal{S}} C\{AJ_n\}^* C\{AJ_n\}^T d\xi d\eta.
\end{eqnarray*}
Because $\mathcal{E}_{\mathcal{S}}$ is always positive, this is a convex quadratic program and thus very efficient global solvers are available given estimates of $d_n$, $B_n$, and $M_n$.  One simple approach is to augment with Lagrange multiplier, $\mu$, and solve the first order optimality condition as a function of $\mu$.  However, as with the other algorithms, we assume only estimates of the field in the image plane are available at each iteration, $C\{Ah_n\}$.  For the Jacobian, we proceed as presented previously by using a constant value obtained assuming $\psi_n = 0$ and ignoring the cross term between DM influence function and aberration,
\begin{equation}
J_0 = \left . \frac{\partial h}{\partial X}\right |_{X_n = 0} = i \frac {2 \pi \lambda_0}{\lambda} F(x,y).
\end{equation}

This greatly reduces the computations needed in the algorithm since the state does not appear in the gradient. The augmented cost function can be written using the lagrange multiplier $\mu$,
\begin{equation}
\mathcal{E}_\mathcal{M} = \frac{1}{2} (X_n + \delta X_{n+1})^T W^{-1} (X_n + \delta X_{n+1} )  +  \mu (\delta X_{n+1}^T M_0 \delta X_{n+1} + B_n \delta X_{n+1} +d_n - 10^{-\mathcal{C}} )
\end{equation}
The optimality condition corresponds to solving for the zeros of the derivative of $\mathcal{E}_\mathcal{M} $,
\begin{equation}
\delta X_{n+1} (\mu) = -(\mu I + 2 W M_0)^{-1} (\frac{1}{\mu} X_n + W B_n^T)
\end{equation}
where,
\begin{eqnarray}
B_n & = & 2\Re \left \{ \int \int_{\mathcal{S}} C\{Ah_n\}^* C\{AJ_0\} d\xi d\eta \right \} = \left<E_f,  i \frac {2 \pi \lambda_0}{\lambda}C\{A F \} \right>_{\mathcal{S}} \\
M_0 & = & \int \int_{\mathcal{S}} C\{AJ_0\}^* C\{AJ_0\}^T d\xi d\eta=  -( \frac {2 \pi \lambda_0}{\lambda})^2\left<C\{A F \} , C\{A F \} \right>_{\mathcal{S}}. 
\end{eqnarray}

\begin{figure}[h]
\begin{center}
\includegraphics[width = 6in]{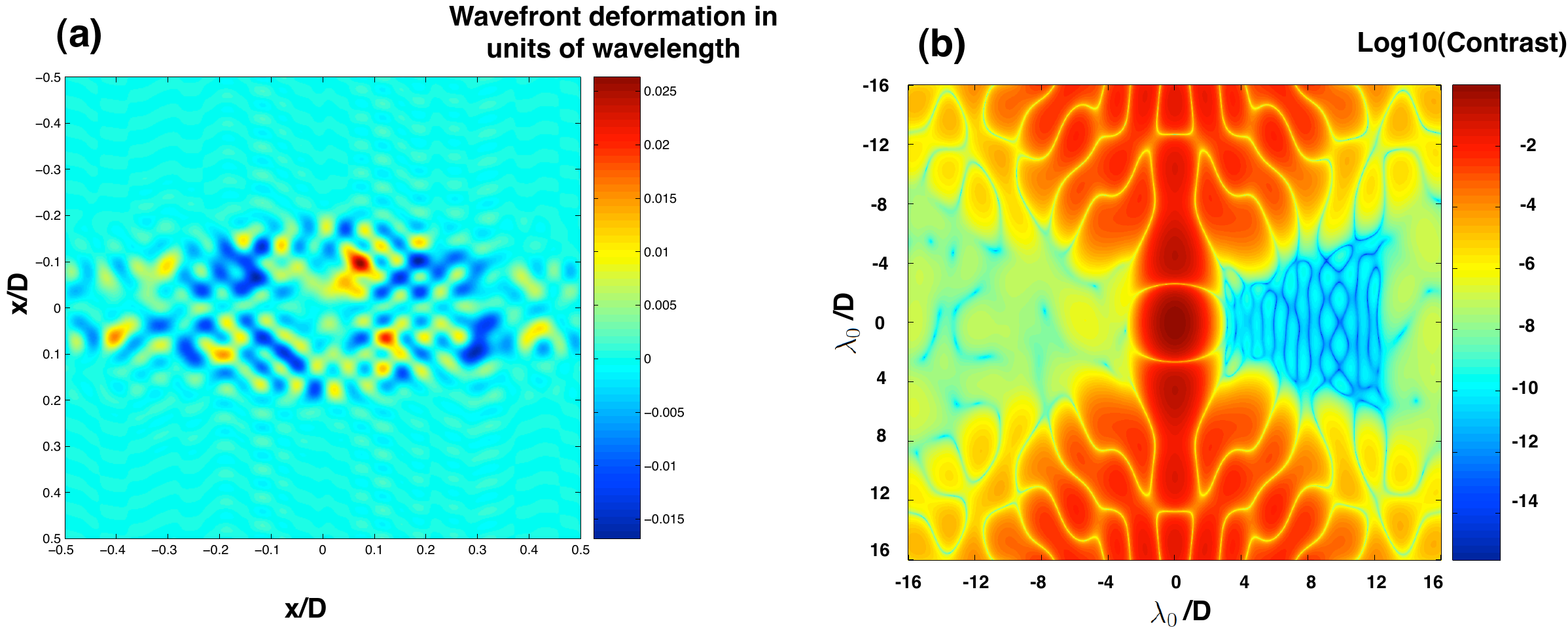}
\end{center}
\caption[Numerical results of the Stroke Minimization algorithm using a $10^{-10}$ Shaped Pupil]{Numerical results of the Stroke Minimization algorith using a $10^{-10}$ Shaped Pupil. Top: DM deformation in radians. Bottom: Log(Corrected PSF).}
\label{figIcorrSrokMinImag}
\end{figure}

\begin{figure}[h]
\begin{center}
    \includegraphics[width = 6in]{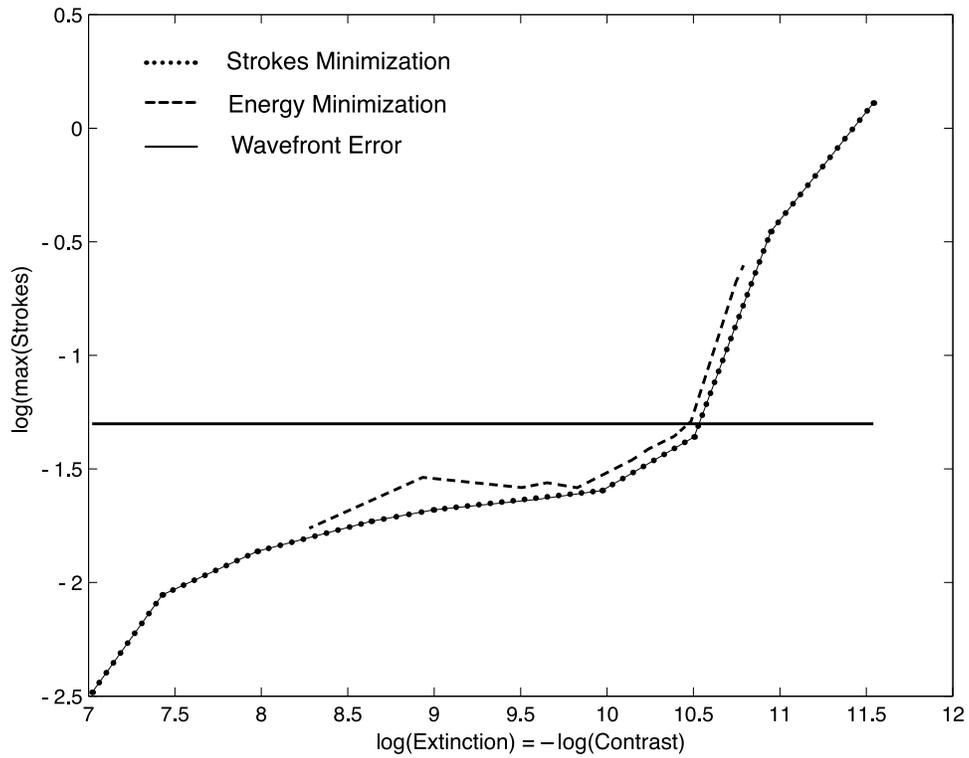}
\end{center}
\caption{Comparison between the peak to valley actuator strokes necessary to correct a given wavefront error using Energy Minimization and Stroke Minimization.}
\label{figLogStrokesPlot}
\end{figure}

In order to find the optimal $\mu^{\star}$, we use a line search that finds the smallest $\mu =  \mu^{\star}$ such that the contrast constraint is satisfied. Indeed, all the $\mu$-dependent terms in the modified cost function correspond to a penalty that weighs the relative importance of contrast with respect  to actuator minimization. Thus, the more stringent the contrast constraint, the larger $\mu^{\star}$ becomes.  Based on this qualitative approach, we start our algorithm with a small $\mu_0$, compute $X^{\star}(\mu_0)$, simulate the propagation of these DM commands through the system,  and increase $\mu$ until the contrast constraint is satisfied. We also begin with a smaller contrast target to ensure small DM changes at each iteration and slowly increase the target contrast until the goal of $10^{-10}$ is reached.  We call this algorithm ``Stroke Minimization,'' since it finds the smallest deformations that achieve a target contrast.
In Fig.~\ref{figIcorrSrokMinImag} we show, via simulation, the PSF that results from one iteration of the Stroke Minimization algorithm at $\lambda = \lambda_0$, with $C_{Target} = 10^{-10}$.  Fig.~\ref{figLogStrokesPlot} shows a comparison between the strokes obtained using a direct minimization of $\mathcal{E}_\mathcal{S}$ and Stroke Minimization, for which the strokes are smaller by a factor of two to five. For  this simulation, the DM was modeled using an influence function basis, similar to the algorithms we use on our testbed. 

\subsection{Experimental results}
\label{subsec:strokemin_expresults}

\subsubsection{High contrast PSFs and convergence curves}
\begin{figure}[h]
\begin{center}
    \includegraphics[width = 6in]{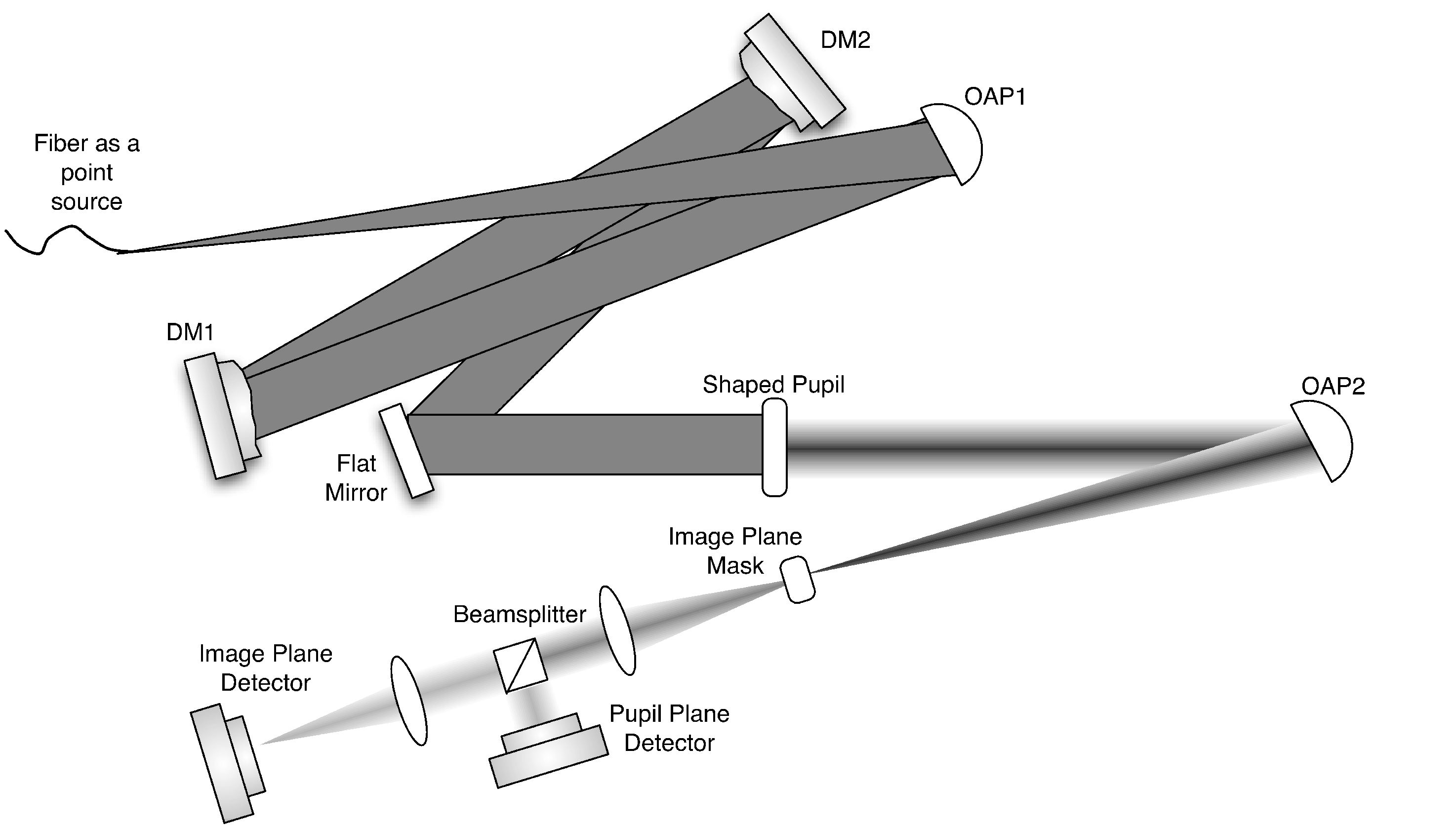}
\end{center}
\caption[Optical layout of the Princeton High Contrast Imaging Laboratory]{Optical layout of the princeton high contrast imaging testbed. For the experiment presented in \S~\ref{subsec:strokemin_expresults}, only one DM is used for wavefront control and only the image plane camera is used for wavefront sensing.}
\label{figHCILSetup}
\end{figure}

\begin{figure}[h]
\begin{center}
    \includegraphics[width = 6in]{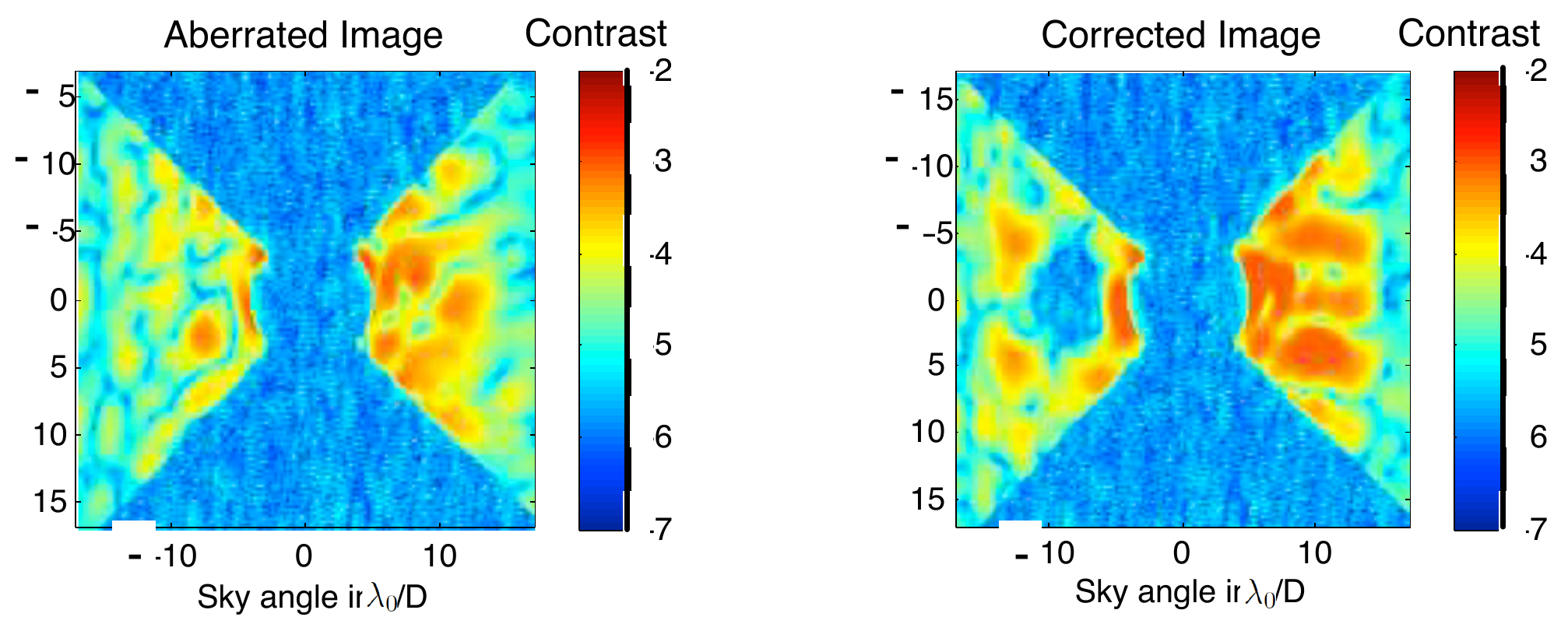}
\end{center}
\caption[Aberrated and corrected PSF on the testbed]{Aberrated (Left) and corrected (Right) PSF on the Princeton testbed in log(contrast). The wavefront is flattened so that half of the image plane exhibits a dark hole over a specified region, in this case X = 7-10 $\lambda/D$ and Y = -2.5-2.5 $\lambda/D$.  This monochromatic experiment used an illumination wavelength of $635$ nm}
\label{figHCILPSFs}
\end{figure}

The optical layout of the Princeton High Contrast Imaging Laboratory is shown in Fig.~\ref{figHCILSetup}. It consists of two six-inch off-axis parabolic mirrors to collimate and refocus the beam, a shaped pupil coronagraph, and two sequential DMs for control.  The DM(s) are 1 cm on a side, and neither  is located in a plane conjugate to the pupil. For the experiment presented here, we use  only the image plane camera and control only one of the two DMs, while no voltage is applied to the other one.  The dark hole obtained using the Stroke Minimization algorithm is shown in Fig.~\ref{figHCILPSFs} and exhibits a dark hole 2 orders of magnitude deeper than the non-corrected one. The dark hour glass shape in the image plane corresponds to a binary mask which suppresses the bright central core and vertical wings of the PSF. Such an image plane mask is used to mitigate the limited dynamic range of our camera.
\subsubsection{Current Contrast Limitations at the Princeton Testbed}

\begin{figure}[h]
\begin{center}
    \includegraphics[width = 5in]{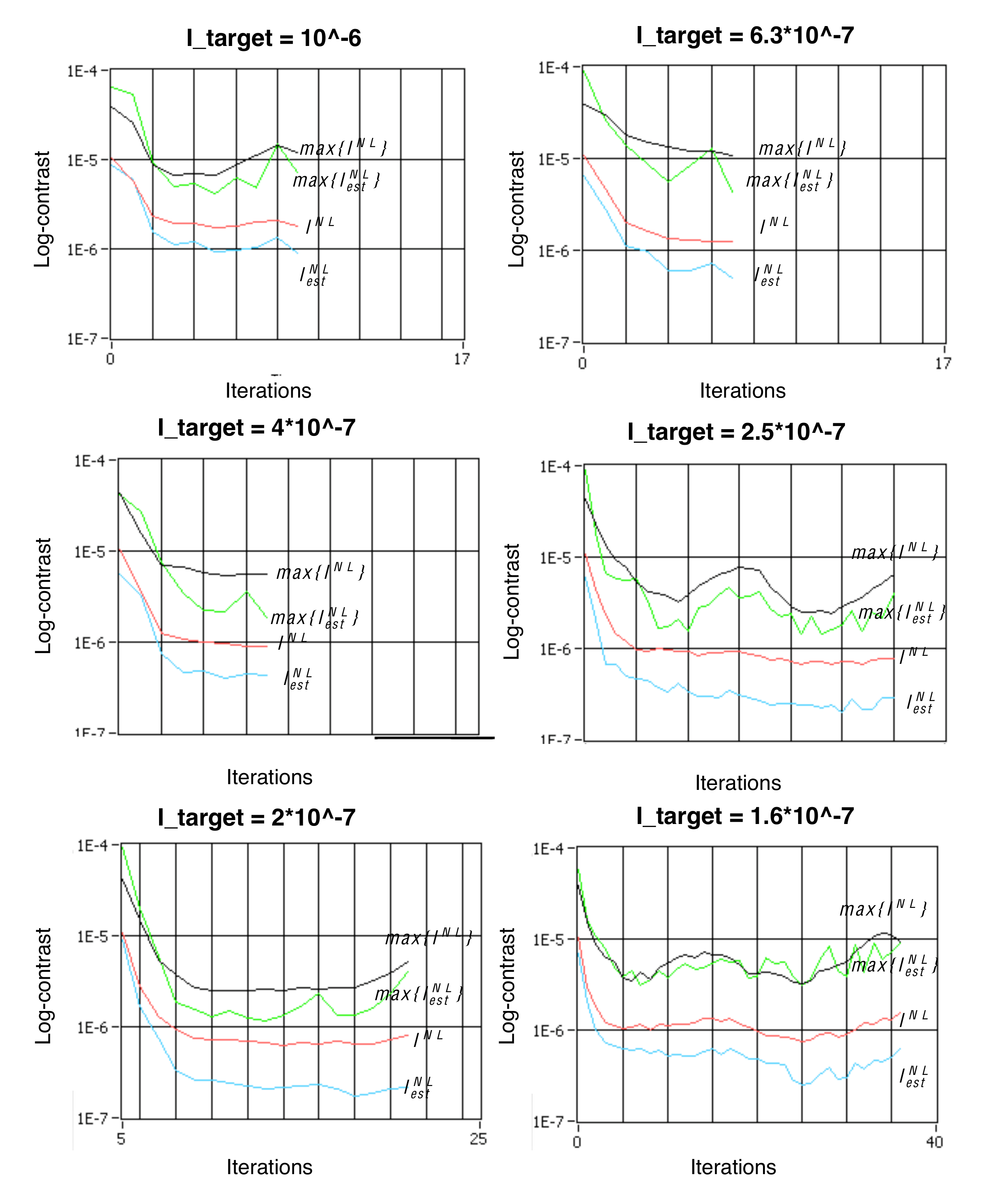}
\end{center}
\caption[Experimental results for six different target contrasts]{Experimental results for six different target contrasts. The top two curves correspond respectively to the maximum of the intensity in the dark hole and the maximum of the estimated intensity in the dark hole. The bottom two correspond to the average intensity in the dark hole and the average estimated intensity in the dark hole. Note that when the algorithm converges the average estimated intensity is equal to the target contrast. Also, note that these results were obtained on the Princeton testbed prior to the installation of the second DM, and therefore, the contrast limit is slightly better than other results shown in this paper.}
\label{figLabResultsStrokeMin}
\end{figure}

Stroke Minimization also proves to be an effective diagnostic tool for the limitations of the testbed.  Fig.~\ref{figLabResultsStrokeMin} shows experimental convergence curves obtained using the Stroke Minimization algorithm.  Each panel illustrates the contrast versus iteration curve for a given target contrast. We use the following notations:
\begin{itemize}
\item $I_{Target}$: Target contrast setting the contrast constraint of the wavefront control algorithm
\item $I_{est}^{NL}$: Integrated magnitude square of the estimated field in the Dark Hole at each iteration. 
\item $I^{NL}$: Integrated intensity actually measured in the Dark Hole at each iteration
\end{itemize}
The last three panels of Fig.~\ref{figLabResultsStrokeMin} illustrates this divergence. This divergence is due to the inherent bias of our DM diversity estimator. We note a discrepancy between the estimated intensity, $I_{est}^{NL} $, and the actual intensity in the dark hole, $I^{NL}$. This implies an estimation bias that can be caused either by incoherent light landing in the dark hole or a systematic error in the estimation scheme. When $I_{Target} = 1.6 \times 10^{-7}$ the iterative loop diverges, primarily due to the large bias in the field estimate. Note that the contrast in the dark hole before correction in Fig.~\ref{figLabResultsStrokeMin} is $10^{-5}$, an order of magnitude better than in Fig.~\ref{fig2DM_results}.  
This estimate bias arises because the estimation algorithm relies upon a linearity assumption of the DM, perfect knowledge of the influence function, and a perfect knowledge of the voltage to deformation transfer function. When applied to the Princeton testbed, these assumptions produce an intrinsic bias in the estimated field. Each iteration of the stroke minimization algorithm minimizes the DM deformation under the constraint that the norm squared of the sum of estimated field and effect of the DM is below $I_{Target}$. If we decompose the estimate as $\delta E_f+ E_f(X_n)$, where $\delta E_f$ is the bias, then the constraint can be written as:

\begin{equation}
\| \delta E_f+ E_f(X_n)+C \{AF \} .  \delta X_k   \|^2 < I_{Target}
\end{equation}
where $ \delta E_f$ is the estimation bias and $\| \; \|^2$ is the norm square. This implies that the constraint in the Stroke Minimization loop is actually
\begin{equation}
 \| \delta E_f  \|^2 +  \|  E_f (X_n) + C \{ AF \}  \delta X_n \|^2 - 2 \Re \left[ \left< \delta E_f, E_f(X_n) + C \{ AF \} .  \delta X_n  \right>_{\mathcal{S}}  \right] < I_{Target}
\end{equation}
If we write the integrated value of the bias as  $I_{bias} =  \| \delta E_f  \|^2 $, the Cauchy Schwartz inequality yields:
\begin{equation}
\|  E_f (X_n) + C \{ AF \}  \delta X_n \|^2 < I_{Target} - I_{bias} + 2 \sqrt{I_{Bias}} \sqrt{\mathcal{E}_\mathcal{S}}
\end{equation}
and the quadratic contrast constraint is more stringent than it is supposed to be.  When $I_{bias}< I_{Target} $ the algorithm is quite insensitive to the estimation error, as seen on the first three panels of Fig.~\ref{figLabResultsStrokeMin}. However, when these two quantities become similar then at some iteration $n_i$  the intensity constraint becomes too stringent; the algorithm seeks to correct for wavefront errors that are larger than the ones actually present in the testbed. This yields a set of deformation coefficients that is too large for the assumptions of the algorithm to be valid. The last three panels of Fig.~\ref{figLabResultsStrokeMin} illustrates this divergence. The upwards trend in the last few iterations indicates that the algorithm might diverge if ran for a few more iterations. When tested the only the last panel actually diverges, for the two other case the algorithm lead to a oscillatory regime in contrast. Solutions to circumvent these limitations include estimation algorithms that do not use the DM as a source of diversity, or adaptive algorithms that build an on-the-fly model of the non-linear response of the DM and include that model in the estimation stage.

\section{Symmetric Dark Hole with two DMs}
\label{sec:two_DMs}

Another value of the stroke minimization algorithm is that it is easily modified to incorporate multiple DMs.  We show in \citet{2007ApJ...666..609P} how multiple DMs are necessary to achieve symmetric dark holes on both sides of the image plane and for achieving high contrast in broader bands. The ability to provide a wavelength independent lever for wavefront correction is the main advantage of controllers based on two sequential DMs since it will ultimately enable broadband observations of exo-planets and thus greatly facilitate their spectral characterization.  The actual implementation of control algorithms in broadband using these methods can be either done using a series of monochromatic estimations, as shown by \citet{2007AAS...21113520G}, or a single wavelength estimation coupled with some priors on the symmetries of  the PSF. In either case, once the estimation is complete the stroke minimization algorithm can be applied in order to retrieve the DM commands. Discussing relative performances of wavefront retrieval methods under polychromatic light is beyond the scope of this paper, and here we chose to only focus on the intricacies of the implementation of a two DMs control algorithm. As a consequence we will only present on monochromatic results, using the same estimator as in \S~\ref{subsec:strokemin_expresults}, using the two DMs to produce a symmetric dark hole.

\subsection{Monochromatic Stroke Minimization with two sequential DMs: general algorithm}
\label{subsec:monostrokemin}
Here we do not delve into to the details of the several single DM correction algorithms presented above and solely focus on Stroke Minimization. To begin, we write the matrix of actuator commands as a concatenation of the coefficients of each DM, $X= [X^{(1)} X^{(2)}]$.   When we consider the case where DM2 is conjugated with the shaped pupil (or the Lyot plane of the coronagraph) and separated from DM1 by a distance $z$, the field at DM2, before the final Fourier Transform, can be written as
\begin{equation}
E_{DM2}(x,y) = A(x,y)h^{(2)}(\gamma_{\lambda},X) 
\end{equation}
where
\begin{equation}
h^{(2)}(\gamma_{\lambda},X) = e^{\alpha_\lambda(x,y) + i\frac{2\pi}{\lambda} \beta_\lambda(x,y)}   \mathcal{F}_z [e^{i\frac{2\pi}{\lambda}\psi^{(1)}(x,y)}] e^{i\frac{2\pi}{\lambda}\psi^{(2)}(x,y)},
\label{Eq:hdeux}
\end{equation}
$\psi^{(1)}$ and $\psi^{(2)}$ stand respectively for the surface of $DM1$ and $DM2$ at the nth iteration, $\mathcal{F}_z $ is the Fresnel propagation between two surfaces separated by a distance $z$, and $A(x,y)$ is the pupil apodisation.  Using the coronagraph operator $C$ notation, the field in the final image plane is then:
\begin{equation}
E_f = C \left[ e^{\alpha_\lambda(x,y) + i\frac{2\pi}{\lambda} \beta_\lambda(x,y)}   \mathcal{F}_z [e^{i\frac{2\pi}{\lambda}\psi^{(1)}(x,y)}] e^{i\frac{2\pi}{\lambda}\psi^{(2)}(x,y)} \right].
\end{equation}
The general form for the intensity resulting from the effects of the two DMs is then
\begin{eqnarray}
\mathcal{E}_\mathcal{S} &\cong& \underbrace{\int \int_{\mathcal{S}} C\{Ah_n^{(2)}\}^*C\{Ah_n^{(2)}\} d\xi d\eta}_{(\mathcal{E}_\mathcal{S})_n} + 
2 \Re \left \{ \int \int_{\mathcal{S}} C\{Ah_n^{(2)}\}^* C\{AJ_n\} \delta X d\xi d\eta \right \} \nonumber \\
& & + \int \int_{\mathcal{S}} \Delta X ^T C\{AJ_n\}^* C\{AJ_n\}^T \delta X d\xi d\eta \nonumber.
\end{eqnarray}
%
%
where, by virtue of the linearity of the operator $C$, 
\begin{equation}
J_n  = \frac{\partial h^{(2)}}{\partial X} |_{X_n}.
\end{equation}
Thus, just as we presented above for the case of a single DM, we are seeking to solve the following optimization problem:
\begin{eqnarray*}
\mbox{minimize} & & \frac{1}{2} \sum_{k=1}^N a_k^2 \\
\mbox{subject to} & & \mathcal{E}_{\mathcal{S}} \le 10^{-\mathcal{C}}
\end{eqnarray*}
We proceed using the same sub-quadratic programming approach and solve at each iteration the following subprogram:
\begin{eqnarray}
\mbox{minimize} & & \frac{1}{2} (X_n + \delta X_{n+1})^T W^{-1} (X_n + \delta X_{n+1} ) \nonumber \\
\mbox{subject to} & & \delta X_{n+1}^T M_n \delta X_{n+1} + B_n \delta X_{n+1} +d_n \le 10^{-\mathcal{C}} \nonumber
\end{eqnarray}
However for the case of multiple DMs we need to reduce the modeling of the wavefront controller in such a way that $M_n$, $B_n$ and $d_n$ can be directly computed either from wavefront estimate or the design parameters of the optical set up. Once again this is done by approximating $J_n$ by a constant value $J_0$, which slows down the convergence rate of the algorithm but circumvents the high computational cost associated with the  evaluation of the sensitivity matrix at each iteration. Next we present the model reduction we implemented for our experimental validation.
\subsection{Monochromatic Stroke Minimization with two sequential DMs: model reduction}
While dividing the problem of finding optimal DM strokes in a a series of quadratic sub-programs is a general method applicable to all types of wavefront control architectures, here we are interested in reducing the sub-program in such a way that:
\begin{itemize}
\item The sensitivity matrix is computed only once, before any correction
\item The sensitivity matrix provides two degrees of freedom to correct on both sides of the image plane
\item The sensitivity matrix provides two degrees of freedom for wavelength independent wavefront errors and those proportional to $1/\lambda$ \cite{2007ApJ...666..609P}
\end{itemize}
We show here how to reduce Eq.~\ref{Eq:hdeux} in such a fashion. The main difference between Eq.~\ref{eq:E0_h}  and Eq.~\ref{Eq:hdeux} is the presence of a Fresnel propagation between the two DMs . This propagation is what provides both the symmetric and wavelength levers. We start by calculating the jacobian of the effect of the DM in the plane where the aberrations are estimated,
\begin{eqnarray}
&&J_n = i \frac{2\pi \lambda_0}{\lambda} e^{\alpha_\lambda(x,y) + i\frac{2\pi}{\lambda} \beta_\lambda(x,y)} \\ & & \left[\begin{array}{cc}  e^{i\frac{2\pi}{\lambda}\psi^{(2)}_n(x,y)} \mathcal{F}_z [e^{i\frac{2\pi}{\lambda}\psi^{(1)}_n(x,y)} F^{(1)}(x,y) ] \; \ ; \mathcal{F}_z [e^{i\frac{2\pi}{\lambda}\psi^{(1)}(x,y)}]  e^{i\frac{2\pi}{\lambda}\psi^{(2)}_n(x,y)}F^{(2)}(x,y)\end{array}\right]\nonumber
\end{eqnarray} 
where $F^{(j)}(x,y)$ is the matrix representing the basis function for the $j$ th DM. We proceed as previously, namely we choose to evaluate this jacobian around $X = 0$ and ignore the cross talk between the DMs and the aberrations. We also ignore the cross talk between the two DMs. These approximations lead to a Jacobian that is not exact but making them only lowers the convergence rate and does not change the final solution. The contribution of the second DM becomes:
\begin{equation}
C\{A(x,y)) i \frac{2\pi \lambda_0}{\lambda} e^{\alpha_\lambda(x,y) + i\frac{2\pi}{\lambda} \beta_\lambda(x,y)} \mathcal{F}_z [e^{i\frac{2\pi}{\lambda}\psi^{(1)}(x,y)}]  e^{i\frac{2\pi}{\lambda}\psi^{(2)}_n(x,y)}F^{(2)}(x,y) \} \simeq  i \frac{2\pi \lambda_0}{\lambda} C\{A F^{(2)}\} 
\end{equation}
In order to calculate the contribution of the first DM, that is not conjugated with the plane of the aberrations, we assume, as it is the case for shaped pupils, that the operator $C$ is a fourier transform. Moreover we work under the angular spectrum approximation, and thus the impact of an out of pupil optics is only to multiply the electrical field in  the image plane by a quadratic phase factor. The contribution of the out-of-pupil plane DM is then:
\begin{eqnarray}
&&C\{i A(x,y) \frac{2\pi \lambda_0}{\lambda} e^{\alpha_\lambda(x,y) + i\frac{2\pi}{\lambda} \beta_\lambda(x,y)} e^{i\frac{2\pi}{\lambda}\psi^{(2)}_n(x,y)} \mathcal{F}_z [e^{i\frac{2\pi}{\lambda}\psi^{(1)}_n(x,y)}] F^{(1)}(x,y) \}\\
&\simeq&  e^{- i \frac{\pi \lambda z}{D^2} (\xi^2+\eta^2)} C\{A F^{(1)}\}  \nonumber
\end{eqnarray}
where  again we have used the fact that the gradient is computed around $X=0$ with no aberration. Thus the reduced and linearized effect of the two DMs becomes
\begin{equation}
C \{A J_n \} \simeq C \{A J_0 \}  = i \frac{2\pi \lambda_0}{\lambda}  \left[ \begin{array}{cc} e^{- i \frac{\pi \lambda z}{D^2} (\xi^2+\eta^2)} C\{A F^{(1)}\}    &  C\{A F^{(2)}\}   \end{array}\right] = [ C \{A J_0^{(1)} \} \; C \{A J_0^{(2)} \} ].
\label{ASGradient}
\end{equation}
In our laboratory we have implemented the stroke minimization algorithm with dual DMs using this reduced Jacobian. Namely, at each iteration we solve the quadratic subprogram
\begin{eqnarray}
\mbox{minimize} & & \frac{1}{2} (X_n + \delta X_{n+1})^T W^{-1} (X_n + \delta X_{n+1} ) \nonumber \\
\mbox{subject to} & & \delta X_{n+1}^T M \delta X_{n+1} + B_n \delta X_{n+1} +d_n \le 10^{-C} \nonumber
\end{eqnarray}
where 
\begin{eqnarray*}
d_n & = &  \left <C\{Ah_n^{(2)}\}^,C\{Ah_n^{(2)}\} \right>_{\mathcal{S}} \\
B_n & = & 2\Re  \left [ \begin{array}{cc} \left < C\{Ah_n^{(2)}\}, C\{A J_0^{(1)}\} \right>_{\mathcal{S}} & \left < C\{Ah_n^{(2)}\}, C\{A J_0^{(2)})\} \right>_{\mathcal{S}}  \end{array} \right ] \\
M & = & \left[\begin{array}{cc}\left < C\{A J_0^{(1)}\}, C\{A J_0^{(1)}\} \right>_{\mathcal{S}} & \left < C\{A J_0^{(1)}\}, C\{A J_0^{(2)}\} \right>_{\mathcal{S}} \\\left < C\{A J_0^{(2)}\}, C\{A J_0^{(1)}\} \right>_{\mathcal{S}} & \left < C\{A J_0^{(2)}\}, C\{A J_0^{(2)}\} \right>_{\mathcal{S}} \end{array}\right]
\end{eqnarray*}
\subsection{Stroke Minimization with two sequential DMs: symmetries and broadband lever}
Note that for small spatial frequencies the angular spectrum is small, $\frac{\pi \lambda z}{D^2} (\xi^2+\eta^2) \ll 1$, resulting in the following first order Taylor expansion
\begin{equation}
\label{grad_smallfreq}
C \{A J_0 \}  \cong \left[ \begin{array}{cc} - \frac{2 \pi^2 \lambda_0 z}{D^2} (\xi^2+\eta^2) C\{A F^{(1)}\} + i \frac{2\pi \lambda_0}{\lambda}  C\{A F^{(1}\}     & i \frac{2\pi \lambda_0}{\lambda}  C\{A F^{(2)}\}   \end{array}\right] 
\end{equation}
If we rearrange the DM commands as $\tilde{X} = [\tilde{X}_1 \; \tilde{X}_2] = [X_1 \; X_1 + X_2]$ then the sensitivity matrix becomes:
\begin{equation}
\label{grad_smallfreq2}
C \{A \tilde{J}_0 \}  \cong \left[ \begin{array}{cc} - \frac{2 \pi^2 \lambda_0 z}{D^2} (\xi^2+\eta^2) C\{A F^{(1)}\} & i \frac{2\pi \lambda_0}{\lambda} ( C\{A F^{(1}\}  + C\{A F^{(2)}\} )   \end{array}\right] 
\end{equation}
Because $F^{(j)}(x,y)$ and $A(x,y)$ are real function in the pupil plane, $C\{A F^{(1)}\} $ and $C\{A F^{(2)}\} $ feature a Hermitian symmetry in the image plane when $C\{ \; \}$ is a fourier transform. Thus the first block of $C \{A \tilde{J}_0 \} $ provides a lever that corrects for Hermitians field distributions in the image plane while the second one provides a lever that corrects for anti-Hermitian distributions. It is the combination of these two independent Hermitian and anti-Hermitian levers that allows us to create symmetric monochromatic dark holes. Note that in the case of a single DM correction there is only an anti-Hermitian lever, thus both sides of the image plane cannot be corrected independently, which constraints the controllable area to only one side of the PSF. These considerations are developed in Appendix \ref{sec:corr1DM} using a phasor representation.\\

Moreover, each of these two independent levers has a different chromatic behavior: the hermitian one scales as $\lambda^0$ and the anti-hermitian one scales at $1 / \lambda$. In terms of wavefront errors this means that a two sequential DM controller can correct over a broadband for what are commonly called amplitude and phase errors in the pupil plane. However, implementing a broadband stroke minimization control algorithm requires a more elaborate estimate of the field in the image plane. One option is to obtain an estimate across the bandpass that provides a set  of $B_n$ and $d_n$ for each wavelength so that the constraint of the quadratic subprogram becomes the integrated broadband intensity. An alternative is to force a monochromatic quadratic subprogram to use the Hermitian lever only to correct for the $\lambda^0$ component of the estimated wavefront and to use the anti-Hermitian lever for its $1/\lambda$ component. While critical for the feasibility of broadband wavefront control and thus the detection of exo-planets, the development and implementation of such estimation algorithms is beyond the scope of this paper and will be presented in a future communication. For the remainder of this article we focus on experimental results that feature a monochromatic symmetric dark hole.

\subsection{Experimental results}
\begin{figure}[h]
\includegraphics[width = 6in]{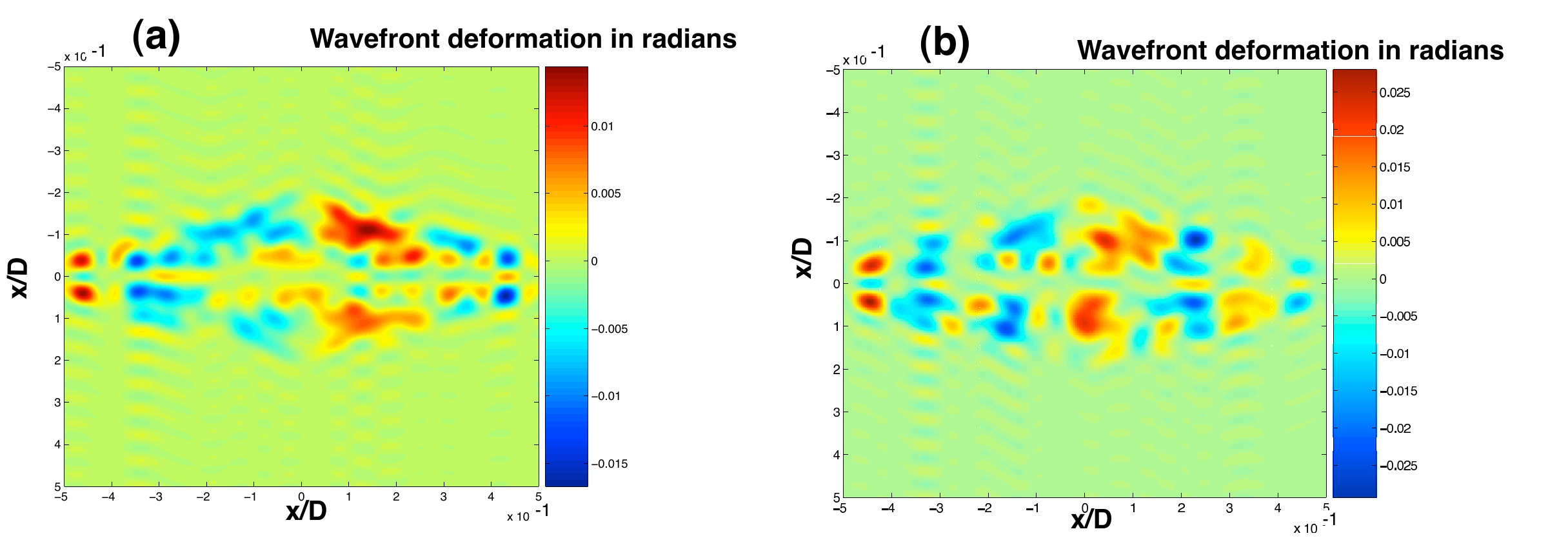}
 \caption[DM surfaces in radians obtained using the two DMs Stroke Minimization algorithm]{DM surfaces in radians obtained using the two DMs Stroke Minimization algorithm. The algorithm used here is designed to operate monochromatically and does not take advantage of the broadband capabilities of the wavefront controller. Pupil size: $D = 3$ cm and DM separation: $z = 1$ m}
\label{figDM12SrokMin2DM10E10}
\end{figure}

\begin{figure}[h]
\includegraphics[width = 6in]{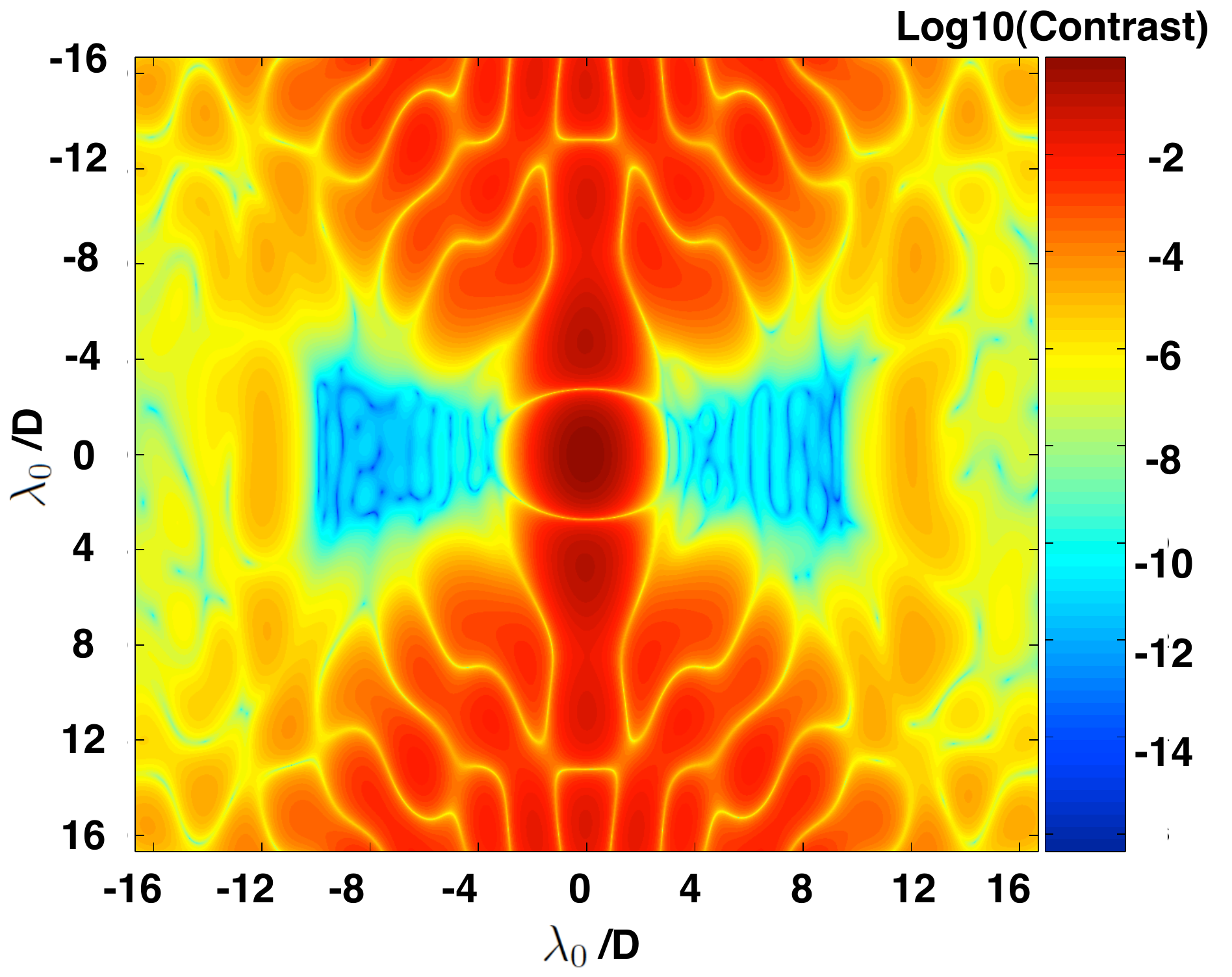}
 \caption[Monochromatic PSF resulting from two DM wavefront correction using a monochromatic Stroke Minimization algorithm]{Monochromatic PSF resulting from two DM wavefront correction using a monchromatic Stroke Minimization algorithm. 
  $D = 3$ cm and $z = 1$ m.}
\label{figIcorrSrokMin2DM10E10plusSYM001Z1}
\end{figure}

\begin{figure}[h]
\begin{centering}
\includegraphics[width = 6in]{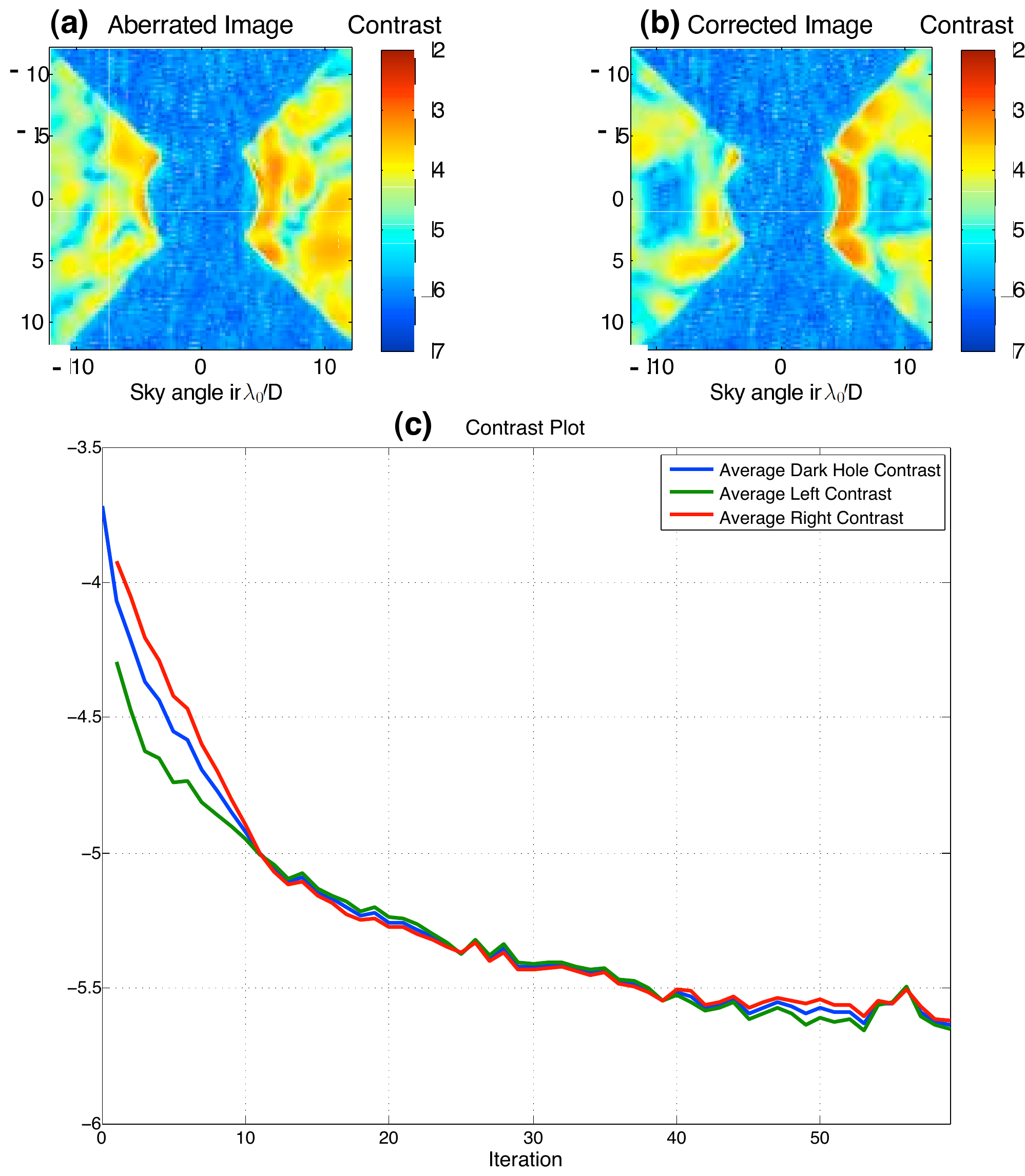}
 \caption{Aberrated image (left) and corrected image(right) of the 2 DM stroke minimization symmetric dark hole experiment. Bottom:  A plot of contrast vs. iteration in each of the two dark holes and in the combination of the two.}
\label{fig2DM_results}
\end{centering}
\end{figure}

Fig.~\ref{figDM12SrokMin2DM10E10} shows the DM surfaces obtained using this algorithm to create the symmetric $10^{-10}$ monochomatic dark hole that is presented on the top panel of Fig.~\ref{figIcorrSrokMin2DM10E10plusSYM001Z1}.  For these numerical simulations we have used square DMs of size $D$ = 3 cm that are separated by $z = 1$ m. 
In this section, we present the first results of a symmetric dark hole using two DMs in sequence to correct for errors on both sides of the image plane.  This experiment was performed in monochromatic 635 nm light using the stroke minimization algorithm as described in \S~\ref{subsec:monostrokemin}.  As before, the estimate of the wavefront was obtained using an algorithm based on the application of diversity on the surface of one of the DMs.  As shown in Fig.~\ref{figHCILSetup}, neither of the two DMs is in a plane conjugate to the shaped pupil.  The propagation from each DM to the pupil plane is taken into account using the angular spectrum approximation, just as shown for the first term of Eq.~\ref{grad_smallfreq}

Figure~\ref{fig2DM_results} shows the aberrated image prior to correction as well as the image after 60 iterations of the stroke minimization correction algorithm.  In addition, the figure shows a contrast plot as a function of iteration.  The Dark Hole is from 7-10 $\lambda/D$ in $x$ and -3 to 3 $\lambda/D$ in y.  The average contrast between the two sides of the image plane before any correction is at $1.2\times 10^{-4}$ with the right side starting out worse than the left side.  After 60 iterations, the contrast on both sides of the image has reached to $2.5\times 10^{-6}$.  This figure shows that the stroke minimization algorithm allowed us to improve the on-axis light extinction by almost two orders of magnitude.
\section{Conclusion}
\label{sec:conclusion}
In this paper we presented a novel general method to solve the non-linear inversion problem associated with the correction of quasi-static wavefront errors. This novel algorithm that we named Stroke Minimization circumvents the dimensionality of the problem and allows a selection of the regularization parameters that is directly related to the target contrast desired in the Dark Hole, where exo-planets are expected to be seen. It can also easily be generalized to multi-DM systems. In this communication, we used this algorithm to accomplish the first experimental proof of a symmetric high contrast PSF obtained using two sequential DMs. This is a significant experimental milestone for the field of high contrast imaging since it not only doubles the search space of coronagraphs, but also proves that amplitude and phase errors can be simultaneously corrected. A full chromatic characterization of this solution will be presented in a subsequent communication.  

\appendix

\section{Physics of wavefront correction}
\label{sec:corr1DM}

\begin{figure}[h]
\begin{center}
\includegraphics[width = 5in]{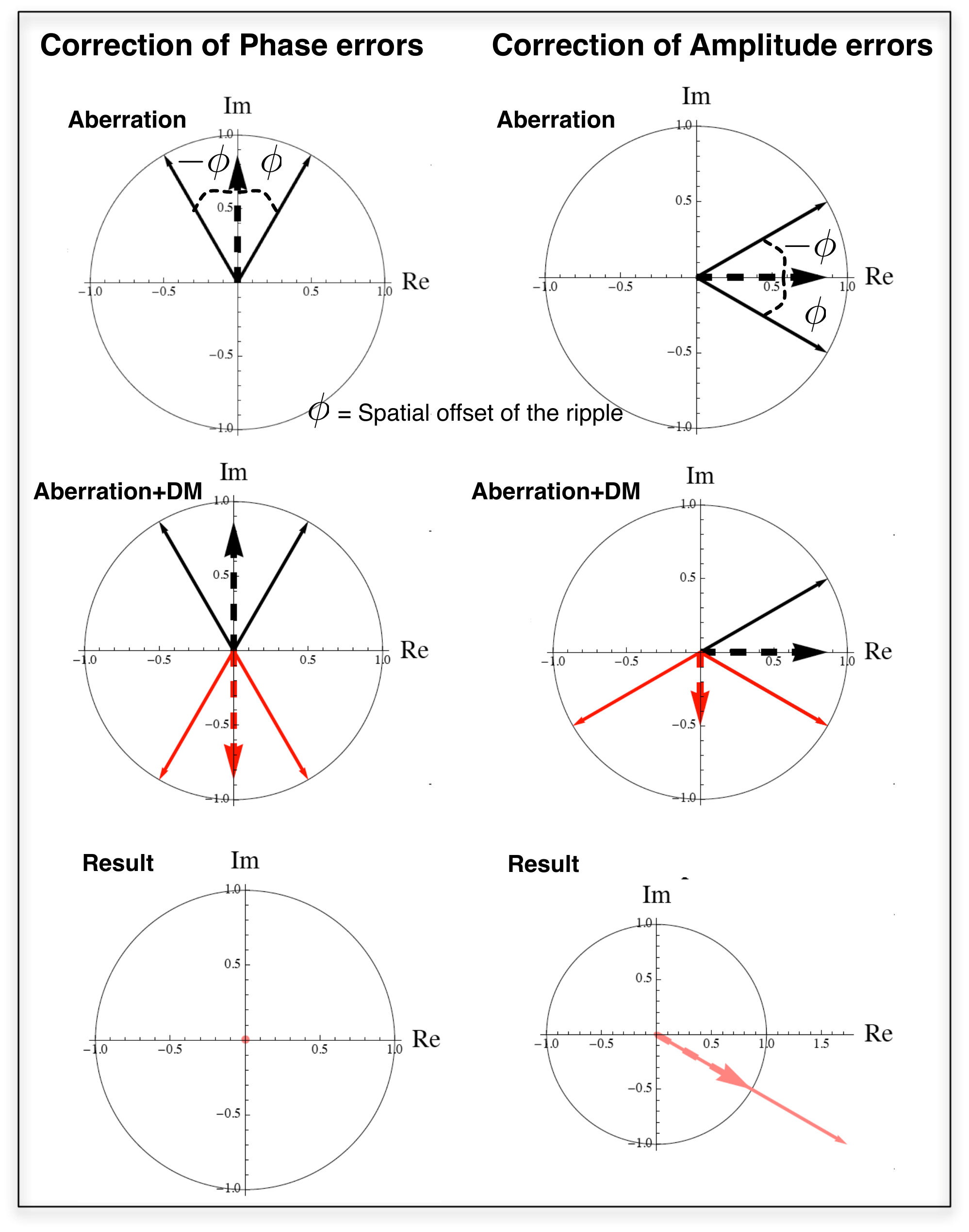}
\end{center}
 \caption{Half dark hole correction using one DM. Left column: a phase aberration can theoretically be compensated at
all wavelengths by a matching DM setting to cancel out the total electric field on both sides of
the optical axis. Right column: In the case of amplitude aberrations, a DM setting can be found that will exactly
compensate the amplitude error on one side of the optical axis, and at a single wavelength}
\label{fig1DMCorrPhasors}
\end{figure}
\begin{figure}[h]
\begin{center}
\includegraphics[width = 5in]{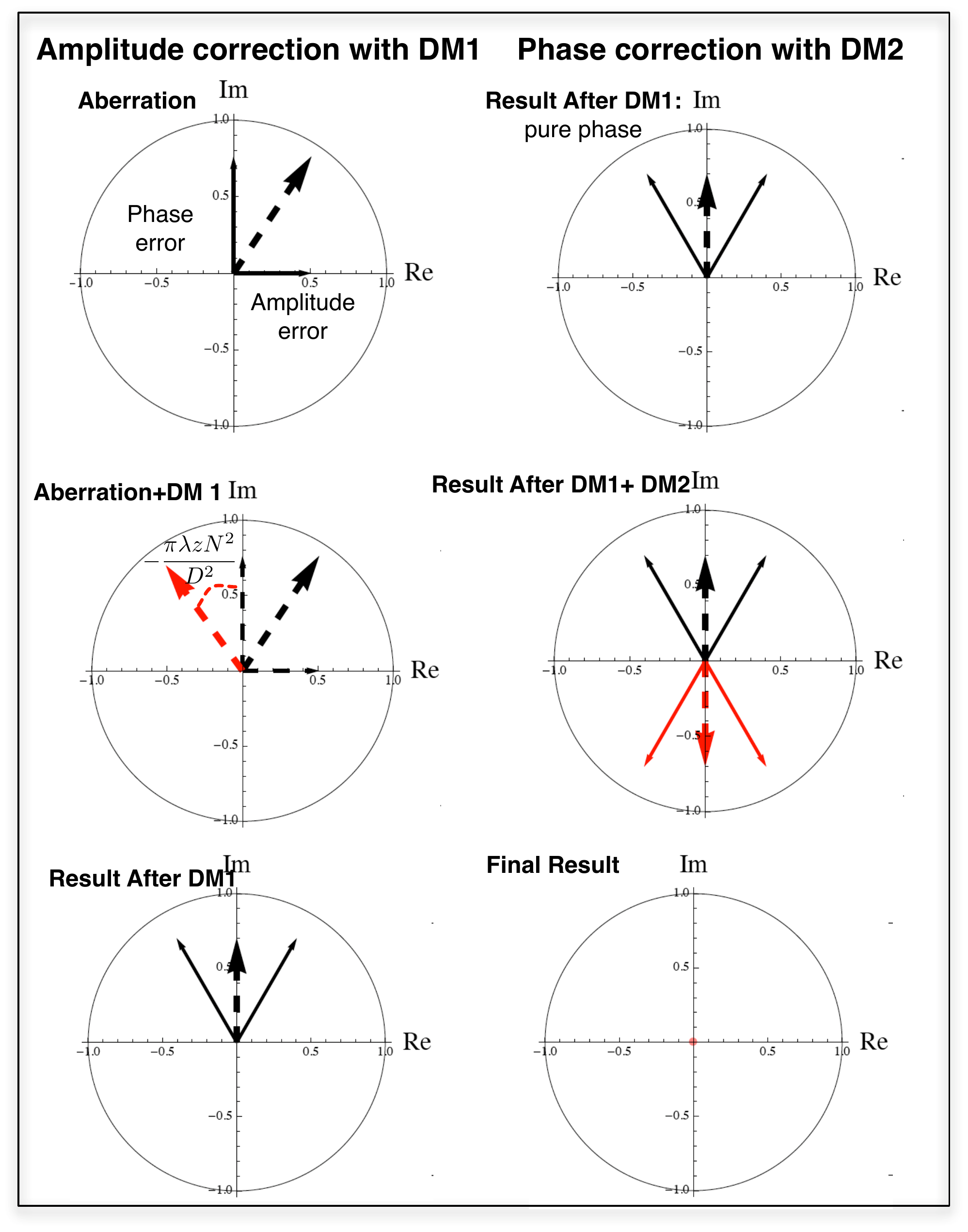}
\end{center}
\caption[Broadband amplitude correction using 2 DMs: complex phasor illustration]{Broadband amplitude correction using 2 DMs: complex phasor illustration. Left Column: one can find a DM1setting that will cancel the amplitude error achromatically on both sides of the optical axis after propagation through the system (equivalent to a phasor rotation in the angular spectrum approximation). Right column: the phase error induced by DM1, together with the phase error accumulated through the system, is then taken out at all wavelengths by thesecond DM, resulting in broad-band 2-sided light cancellation.}
\label{figVectorCorrection2DM}
\end{figure}

We present here a qualitative explanation of the physics involved in a single-DM wavefront compensator. We earlier mentioned that such a device could only create a dark hole in half of the image plane (c.f., \citet{BrownBurrows}), and we show here how it achieves such a feature. First, consider a phase error in the pupil plane that is composed of only one harmonic component:
\begin{equation}
E_{abb}^{Pup}(x,y) = e^{i \frac{\lambda_0}{\lambda} \cos\left(\frac{2 \pi}{D} (mx+ny) + \phi \right)}
\end{equation}
Then, it can be corrected using a deformable mirror to conjugate the field in the pupil plane:
\begin{equation}
E_{DM}^{Pup}(x,y) = e^{- i \frac{\lambda_0}{\lambda} \cos \left( \frac{2 \pi}{D} (mx+ny) + \phi \right)}
\label{EqCorr1DMPhys}
\end{equation}
This correction is perfect for all the wavelengths.
Now consider the case of an amplitude error: 
\begin{equation}
E_{abb}^{Pup}(x,y) =\cos\left(\frac{2 \pi}{D} (mx+ny) + \phi \right)
\end{equation}
and of a DM surface that is such that, to first order:
\begin{equation}
E_{DM}^{Pup}(x,y) = - i \frac{\lambda_0}{\lambda} \sin \left( \frac{2 \pi}{D} (mx+ny) + \phi \right)
\end{equation} 
Then the residual field in the pupil plane is:
\begin{equation}
E_{Res}^{Pup}(x,y) =\frac{1}{2}  (1 -\frac{\lambda}{\lambda_0}) e^{i \left(\frac{2 \pi}{D} (mx+ny) + \phi \right)} + \frac{1}{2}  (1 + \frac{\lambda}{\lambda_0}) e^{-i \left(\frac{2 \pi}{D} (mx+ny) + \phi \right)}
\end{equation}
When $\lambda = \lambda_0$ then the positive spatial frequencies are corrected while the negative ones are not compensated. We illustrate this feature using a phasor representation. Fig.~\ref{fig1DMCorrPhasors} shows how the spatial variations of amplitude and phase errors are represented as pulsating phasors that can be decomposed into two rotating phasors in the complex plane. The clockwise rotating phasor corresponds to the contribution of the ripple in the right half of the image plane and the anti-clockwise in the left half of the image plane. One can cancel the clockwise component of an amplitude error using the anti-clockwise component of a phase deformation introduced by a DM.  This concept is illustrated in Fig.~\ref{fig1DMCorrPhasors}. This is exactly the approach carried out in the ``speckle-nulling'' algorithm (ref Borde Traub), where there is no wavefront estimation and the alignment of the phasor occurs via a trial and error process which considerably lengthens the convergence time (See \S~\ref{sec:introduction}).  
A similar analysis can be performed for wavefront correction using 2 DMs showing that the wavefront actuator can correct both amplitude and phase aberrations under a broadband illumination.  This capability was first shown in \citet{Stuart2DMsSPIE}.  Using the set up of Fig.~\ref{figHCILSetup}, where DM2 is in a plane conjugate to the final imaging lens. DM2 can be used to correct phase errors as shown by Eq.~\ref{EqCorr1DMPhys}.  Amplitude errors can then be compensated using a combination of DM1 and DM2.  Assume the amplitude aberration, located at DM2, is such that:
\begin{equation}
E_{pup,abb}(x,y) =\cos\left(\frac{2 \pi}{D} (mx+ny) + \phi \right)
\end{equation}
We choose the surface of DM1, such that the linear contribution of DM1 to the field is:
\begin{equation}
E_{DM1,pup}(x,y) = i \frac{\lambda_0}{\lambda}  \frac{D^2}{\pi z (n^2+m^2)} \cos \left( \frac{2 \pi}{D} (mx+ny) + \phi \right)
\end{equation}
Then, using the results of \cite{2007ApJ...666..609P} and \cite{2006ApOpt..45.5143S}, the propagation from DM1 to DM2 of this field is given by:
\begin{equation}
E_{DM1,pup}(x,y) = - i \frac{\lambda_0}{\lambda}  \frac{D^2}{\pi z \lambda_0 (n^2+m^2)} e^{-i \frac{\pi \lambda z (n^2+m^2)}{D^2}}\cos \left( \frac{2 \pi}{D} (mx+ny) + \phi \right)
\end{equation}
We work in the low-to-mid spatial frequency regime, so here again we can assume that $\frac{\pi \lambda z (n^2+m^2)}{D^2} \ll 1$. With this first order approximation of the angular spectrum factor the contribution of DM1 at DM2 becomes:
\begin{equation}
E_{DM1,pup}(x,y) = - i \frac{\lambda_0}{\lambda}  \frac{D^2}{\pi z \lambda_0 (n^2+m^2)} \cos \left( \frac{2 \pi}{D} (mx+ny) + \phi \right) - \cos \left( \frac{2 \pi}{D} (mx+ny) + \phi \right)
\end{equation}
Therefore choosing:
\begin{equation}
E_{DM2,pup}(x,y) =  i \frac{\lambda_0}{\lambda}  \frac{D^2}{\pi z \lambda_0 (n^2+m^2)} \cos \left( \frac{2 \pi}{D} (mx+ny) + \phi \right) 
\end{equation}
yields a broadband cancellation of the amplitude error. These considerations are illustrated using a complex phasor representation on Fig.~\ref{figVectorCorrection2DM}, where DM1 corrects for the amplitude errors and DM2 for the phase errors. Note that there is an important assumption underlying this result: the small angular spectrum regime is required in order to obtain the broadband property of the two DMs controller. This yields an outer working angle limit, dependent on the optical design of the controller, that has been derived in \citet{2007ApJ...666..609P}.
%

\section*{Acknowledgements}
The research described in this publication was carried out at the Jet Propulsion Laboratory, California Institute of Technology, under a contract with the National Aeronautics and Space Administration. The first author was supported by an appointment to the NASA Postdoctoral Program at the JPL, Caltech, administered by Oak Ridge Associated Universities through a contract with
NASA.

\section*{Figures}

\newpage

\newpage

\newpage

\newpage

\newpage

\newpage

\newpage

\newpage

\newpage

\newpage

\end{document}